\documentclass[11pt]{article}

\columnwidth\textwidth

\begin{document}

\title{Comment on 'Another form of the Klein-Gordon equation'}

\author{Andreas Aste\\ 
Institut f\"ur Theoretische Physik der Universit\"at Basel,
Klingelbergstrasse 82, 4056 Basel\\
Switzerland}

\date{August 9, 2001}

\maketitle
 

\begin{abstract}
It is shown that the alternative Klein-Gordon equation with
positive definite probability density proposed in a letter by
M.D. Kostin does not meet the requirements of relativistic (quantum) field
theory and therefore does not allow for a meaningful physical
interpretation.
\end{abstract}

\newpage

The alternative formulation of the Klein-Gordon equation \cite{Klein,Gordon}
proposed by M.D. Kostin reads \cite{Kostin}
\begin{equation}
i \hbar \frac{\partial \phi}{\partial t}=+mc^2 \phi+c (\hat{\vec{p}} \vec{\psi})
\label{KG1}
\end{equation}
\begin{equation}
i \hbar \frac{\partial \vec{\psi}}{\partial t}=-mc^2 \vec{\psi}+c
\hat{\vec{p}}\phi
\quad , \label{KG2}
\end{equation}
where $\phi(\vec{r},t)$ and $\vec{\psi}(\vec{r},t)$ are 'scalar' and 'vector' 
probability
amplitudes, respectively, and $\hat{\vec{p}}=-i \hbar \vec{\nabla}$.
Defining the probability density 
\begin{equation}
P=\phi^*\phi+(\vec{\psi}^*\vec{\psi}) \label{KG3}
\end{equation}
and the probability current density
\begin{equation}
\vec{S}=c(\phi^*\vec{\psi}+\phi\vec{\psi}^*) \label{KG4} \quad ,
\end{equation}
one readily derives the probability conservation equation
\begin{equation}
\frac{\partial P}{\partial t}+ \vec{\nabla} \vec{S}=0 \label{KG5} \quad .
\end{equation}
It is a nice feature of the probability density $P(\vec{r},t)$ to be
positive definite, although is is clear that the non-existence
of a posititve definite probability density for the Klein-Gordon
equation is no more a problem in quantum field theory.

Multiplying (\ref{KG1}) with
$(i \hbar \frac{\partial}{\partial t}+mc^2)$
and (\ref{KG2}) by $c\hat{\vec{p}}$ and combining
the results, one obtains
\begin{equation}
\hbar^2 \frac{\partial^2}{\partial t^2} \phi-
c^2 \hbar^2 \vec{\nabla}^2 \phi + m^2c^4 \phi
=0 \quad ,
\end{equation}
i.e. $\phi$ satisfies the Klein-Gordon equation,
but in a similar way one immediately sees that
the components of $\vec{\psi}$ fulfil the (non-covariant) equation
\begin{equation}
\hbar^2 \frac{\partial^2}{\partial t^2} \vec{\psi}-
c^2 \hbar^2 \vec{\nabla}(\vec{\nabla}\vec{\psi}) + m^2c^4 \vec{\psi}
=0 \quad .
\end{equation}
Although the problematic nature of equations (\ref{KG1})-(\ref{KG5})
can be uncovered
easily, their tempting form sometimes leads to confusion and the
equations have even
found their way into literature \cite{Dvoeglazov}. Furthermore,
when the scalar particle described by $(\phi,\vec{\psi})$
is coupled to an electromagnetic potential, different results
are obtained as in the case of the Klein-Gordon equation.
One must therefore ask
if the proposed equations should be treated on an equal footing with the
usual Klein-Gordon equation.

We give simple arguments in the following which show that
the alternative form of the
Klein-Gordon equation is hard to interpret in a meaningful
way.
Obviously, (\ref{KG1}) and (\ref{KG2}) can be cast in a Dirac-like form
\begin{equation}
i \hbar \frac{\partial}{\partial t} \Psi=mc^2 \beta \Psi+c (\vec{\alpha}
\vec{p}) \Psi \quad ,
\end{equation}
with appropriate matrices $\beta$ and $\vec{\alpha}$ and the
four-component wave function
\begin{equation}
\Psi={\phi \choose \vec{\psi}} \quad ,
\end{equation}
or, using a more compact notation in the following where $\hbar=c=1$
\begin{equation}
\{ i \gamma^\mu \partial_\mu-m \} \Psi(x)=\{ \gamma^\mu \hat{P}_\mu-m
\} \Psi(x)=0\quad .
\end{equation}
Then it is easy to show by straightforward calculation
that matrices $S(\Lambda)$ which relate
the wave functions in different coordinates $x, x'$
\begin{equation}
{x'}^\mu = \Lambda^\mu_{\, \, \nu} x^\nu \quad , \quad x^\nu=(ct,\vec{r}) \quad 
,
\quad \Lambda^\mu_{\, \, \rho} \gamma^\rho=S^{-1}(\Lambda)
\gamma^\mu S(\Lambda) \quad ,
\end{equation}
according to
\begin{equation}
\Psi'(x')=S(\Lambda)\Psi(x)=S(\Lambda)\Psi(\Lambda^{-1}x') \quad ,
\end{equation}
\begin{equation}
\{ \gamma^\mu \hat{P}_\mu-m
\} \Psi(x)=\{ \gamma^\mu \hat{P'}_\mu-m\} \Psi'(x')=0
\end{equation}
exist trivially for rotations, but not for
general Lorentz transformations \cite{Hencken}.

A severe problem arises when one considers the propagators for
the proposed theory.
The Dirac equation can be written in an explicit form as follows
\begin{equation}
\left( \begin{array}{cccc}
\hat{p}_0-m & 0 & \hat{p}_3 & \hat{p}_1-i\hat{p}_2 \\
0 & \hat{p}_0-m & \hat{p}_1+i \hat{p}_2 & -\hat{p}_3 \\
-\hat{p}_3 & -\hat{p}_1+i \hat{p}_2 & -\hat{p}_0-m & 0 \\
-\hat{p}_1-i \hat{p}_2 & \hat{p}_3 & 0 & -\hat{p}_0-m
\end{array} \right) \left( \begin{array}{cccc} 
\Psi_1 \\ \Psi_2 \\ \Psi_3 \\ \Psi_4 \end{array} \right)_D=0 \quad
\label{Dirac} ,
\end{equation}
and by formal inversion of the matrix in (\ref{Dirac}) the retarded (advanced)
propagator can be constructed in momentum space
\begin{equation}
\tilde{S}_{R,A} (p) \sim \frac{\gamma^\mu p_\mu+m}{p^2-m^2 \pm {ip_0} 0}
\end{equation}
which has causal support in real space
\begin{equation}
\mbox{supp} \, (S_{R,A}(x)) \subseteq V^{\pm} \quad , \label{support}
\end{equation}
\begin{equation}
V^+=\{x \in {\bf{R}}^4 | x^2 \geq 0, x^0 \geq 0 \} \quad , \quad
V^-=\{x \in {\bf{R}}^4 | x^2 \geq 0, x^0 \leq 0 \} \quad ,
\end{equation}
a fact which expresses,
roughly speaking, the causal structure of the theory \cite{Epstein}.
The support property (\ref{support}) of the tempered distributions
$S_{R,A} \in {\cal{S}}'({\bf{R}}^4)$ means that the product
$<S_{R,A} | f>$ vanishes for all rapidly decreasing test functions
in Schwartz space $f \in {\cal{S}}({\bf{R}}^4)$ which have their
support outside the forward (backward) light cone.
But in the present case, inversion of the differential operator 
\begin{equation}
\left(
\begin{array}{cccc}
\hat{p}_0-m  & -\hat{p}_1 & -\hat{p}_2 & -\hat{p}_3 \\
\hat{p}_1 & -\hat{p}_0-m & 0 & 0 \\
\hat{p}_2 & 0 & -\hat{p}_0-m & 0 \\
\hat{p}_3 & 0 & 0 & -\hat{p}_0-m
\end{array} \right)  
\end{equation}
leads to a result
\begin{equation}
\sim \frac{1}{p^2-m^2}
\left( \begin{array}{cccc}
p_0-m  & -p_1 & -p_2 & -p_3 \\
p_1 &\frac{-p_0^2+p_2^2+p_3^2+m^2}{p_0+m} &
-\frac{p_1 p_2}{p_0+m} & 
-\frac{p_1 p_3}{p_0+m}\\
p_2 & -\frac{p_1 p_2}{p_0+m}
 & \frac{-p_0^2+p_1^2+p_3^2+m^2}{p_0+m} & 
-\frac{p_2 p_3}{p_0+m}  \\
p_3 & -\frac{p_1 p_3}{p_0+m}&
\frac{-p_2 p_3}{p_0+m}
& \frac{-p_0^2+p_1^2+p_2^2+m^2}{p_0+m}
\end{array} \right)  \quad ,
\end{equation}
which is in conflict with the requirements of the local structure
of quantum field theory
due to the non-local operator $\sim (\hat{p}_0+m)^{-1}$ in the propagator.
The description of a scalar particle in the Duffin-Kemmer-Petiau formalism 
\cite{Duffin,Kemmer,Petiau} by
a five-component wave function is equivalent (at least on the classical
level) to the usual Klein-Gordon equation and causes no problems of
that kind \cite{Lunardi}.

\end{document}